\documentclass[12pt,preprint]{aastex}

\usepackage{emulateapj5,apjfonts}
\usepackage{graphics}
\usepackage{amssymb}
\usepackage{onecolfloat}

\lefthead{Heitmann, Higdon, Nakhleh, Habib}
\righthead{Cosmic Calibration}

\begin{document}

\def\head{
\vbox to 0pt{\vss \hbox to 0pt{\hskip 440pt\rm LA-UR-06-2320\hss}
  \vskip 25pt}

\title{Cosmic Calibration}
\author{Katrin Heitmann\altaffilmark{1},
        David Higdon\altaffilmark{2},          
        Charles Nakhleh\altaffilmark{3}, and
        Salman Habib\altaffilmark{4}}
\affil{$^1$ ISR-1, MS D466, Los Alamos National Laboratory, Los
Alamos, NM 87545, heitmann@lanl.gov} 
\affil{$^2$ D-1, MS F600, Los
Alamos National Laboratory, Los Alamos, NM 87545, dhigdon@lanl.gov} 
\affil{$^3$ X-2, MS T087, Los Alamos National Laboratory, Los
Alamos, NM 87545, cnakhleh@lanl.gov} 
\affil{$^4$ T-8, MS B285, Los Alamos National Laboratory, Los
Alamos, NM 87545, habib@lanl.gov} 

\begin{abstract}

The complexity and accuracy of current and future ``precision
cosmology'' observational campaigns has made it essential to develop
an efficient technique for directly combining simulation and
observational datasets to determine cosmological and model parameters;
a procedure we term {\em calibration}. Once a satisfactory calibration
of the underlying cosmological model is achieved, independent
predictions for new observations become possible. For this procedure
to be effective, robust characterization of the uncertainty in the
calibration process is highly desirable. In this {\em Letter}, we
describe a statistical methodology which can achieve both of these
goals. An application example based around dark matter structure
formation simulations and a synthetic mass power spectrum dataset is
used to demonstrate the approach.

\end{abstract}

\keywords{Cosmology: cosmological parameters --- cosmology: theory}}

\twocolumn[\head]

\section{Introduction}

It is widely recognized that, beginning in the last decade, a
transition to an era of ``precision cosmology'' is well underway.
Ongoing and upcoming surveys such as the Wilkinson Microwave
Anisotropy Probe (WMAP, Sper\-gel et al. 2006), the Sloan Digital Sky
Survey (SDSS, Adelman-McCarthy et al. 2006), Planck, the Dark Energy Survey
(DES), the Joint Dark Energy Mission (JDEM), the Large Synoptic Survey
Telescope (LSST), and Pan-STARRS constitute superb sources of
cosmological statistics. These sources include (galaxy, cluster, and
mass) power spectra and cluster mass functions, from which roughly 25
cosmological parameters have to be constrained (see, e.g., Spergel et
al. 2006, Tegmark et al. 2003, Abazajian et al. 2005). The promised
accuracy from future observations is remarkable, as some parameters
can be measured at the 1\% level or better, posing a major challenge
to cosmological theory. Predictions and analysis methods must at least
match -- and preferably substantially exceed -- the observational
accuracy. For many observables, this can only be achieved by
simulations incorporating physical effects beyond the reach of
analytic modeling.

Cosmological simulations already play a key role in the design and
interpretation of observations. Controlling systematics is a necessary
first step, followed by combining simulations with observations to
extract cosmological and model parameters. This cannot be accomplished
by brute force. For example, if every parameter is sampled only
ten times in a twenty-dimensional parameter space, it would require
$10^{20}$ large-scale simulations, which is currently -- and
in the near-term -- quite impossible. Even the variation of only a
subset of the parameters over a sufficient range is infeasible. The
need to develop and employ reliable statistical methods to determine
and constrain parameters robustly is therefore manifest.

In this {\em Letter} we describe a statistical framework to determine
cosmological and model parameters and associated uncertainties from
simulations and observational data (for an overview of the basic ideas
see, e.g., Kennedy \& O'Hagan 2001 and Goldstein \& Rougier 2004). The
framework integrates a set of interlocking procedures: (i) simulation
design -- the determination of the parameter settings at which to
carry out the simulations; (ii) emulation -- given simulation output
at the input parameter settings, how to estimate the output at new,
untried settings; (iii) uncertainty and sensitivity analysis --
determining the variations in simulation output due to uncertainty or
changes in the input parameters; (iv) calibration -- combining
observations (with known errors) and simulations to estimate parameter
values consistent with the observations, including the associated
uncertainty; (v) prediction -- using the calibrated simulator to
predict new cosmological results with a set of uncertainty bounds.

For concreteness, we discuss the framework methodology in terms of a
simple example application: Estimation of five parameters from dark
matter structure formation simulations and a synthetic set of ``WMAP +
SDSS'' measurements of the matter power spectrum. A detailed
description will be provided elsewhere~(S.~Habib et al. in
preparation).

\section{The Statistical Framework}

We employ a Bayesian framework to update prior probability
distributions on cosmological parameters given observational
data. Denoting these parameters collectively by $\theta$ and the
observed power spectrum data by a vector $y_{\rm obs}$, we model the
data as:
\begin{equation}
y_{\rm obs} = \eta(\theta) +\epsilon,
\end{equation}
where $\eta(\theta)$ denotes the simulation output at input setting
$\theta$, and $\epsilon \sim N(0,\Sigma_y)$
where $\Sigma_y$ describes the error structure of the
observations and any potential systematic differences
between the simulated and observed data.

Standard Bayesian estimation (Jeffreys 1961) proceeds using the
likelihood  
\begin{equation}
L(y_{\rm obs}|\theta) \propto |\Sigma_y|^{-1/2}
{\exp}\{-\frac{1}{2}[y_{\rm obs} - \eta(\theta)]^T \Sigma_y^{-1} [y_{\rm
obs} - \eta(\theta)]\},
\end{equation} 
and a prior $\pi(\theta)$ to form the posterior distribution on
$\theta$: 
\begin{equation}
\pi(\theta|y_{\rm obs}) \propto L(y_{\rm obs}|\theta)\pi(\theta).
\end{equation}
Because the resulting posterior distributions are not in any easily
recognized closed form, they must be explored numerically, usually
using Markov Chain Monte Carlo (MCMC) techniques (Besag et al. 1995).
This procedure requires running the (potentially very expensive)
simulation codes many thousands of times as the $\theta$-space is
explored. However, only a limited number ($\sim 100?\, \sim 1000$?) of
runs may be feasible.  Therefore, efficiently combining Bayesian
methods with simulations requires a representation of the code output
(emulator) that can be sampled many thousands of times during the
course of the MCMC in lieu of running the actual code. When queried at
an input setting where a code run is available, the emulator should
reproduce the output of the code.  At other input settings, the
emulator effectively interpolates nearby code runs while including
uncertainty due to the lack of complete knowledge of the code output.
The selection of input settings in the simulation design must be
sufficiently dense that the emulator can accurately mimic the code
output, and also be sufficiently sparse that the simulation campaign
is computationally tractable.

Systematic design of simulation procedures is reviewed in Santner et
al. (2003). We use orthogonal array-based Latin hypercube
sampling~\cite{tang} to fix 128 input settings over the five
parameters. This approach takes an orthogonal array design -- which
ensures that all lower-dimensional projections have desirable
space-filling properties -- and modifies it so that it is also a Latin
hypercube, the most efficient stratified sampling strategy.

The code output for the $i$th input setting is a power spectrum,
$y^{(i)}(k)=\eta(\theta_i)$, viewed as a column vector over the $n_k$
points in $k$ space.  Each of the resulting $n_s = 128$ output spectra
is loaded into a single $n_k \times n_s$ matrix: $y_{\rm sims} =
[y^{(1)}|y^{(2)}|\ldots|y^{(n_s)}].$ This matrix is then subjected to
a singular value decomposition (SVD) to find an efficient empirical
orthogonal representation of the simulation outputs: $[y_{\rm
sims}]_{ij} = [USV^T]_{ij} = \sum_{p=1}^{n_s} \lambda_p [\alpha_p]_i
w_p(\theta_j)$ where the $\alpha_p$'s are $n_k \times 1$ orthogonal
basis vectors in the simulation output space (columns of $U$), the
$\lambda_i$'s are the singular values of the simulation matrix, and
each principal component (PC) weight $w_p(\theta)$ is a $1 \times n_s$
row vector in the parameter space (columns of $V$).  Usually the first
few singular values dominate the remainder, allowing us to keep only a
few of the principal components in the analysis. In what follows, we
have kept five PC's.

The SVD gives the PC weights at the design input settings $(\theta_1,
\theta_2,\cdots,\theta_{n_s})$.  However, in the course of the MCMC,
we need the PC weights at intermediate input settings. We construct
the emulator by putting a spatial Gaussian Process (GP) model on each
PC weight (Sacks et al. 1989, MacKay 1998), a nonlinear interpolation
scheme that works directly on the space of functions. This allows the
emulator to smoothly interpolate the predicted code output between the
design settings, giving an efficient probabilistic representation of
the prediction uncertainty. The spatial parameters controlling the GP
on each component weight are estimated in the course of the MCMC,
thereby constructing the emulator as needed during the calibration
analysis. Details of the procedures used in our code are being
reported elsewhere in the literature.  For recent examples of these
techniques used in practice, see Higdon et al. (2004).

\section{Parameter Estimation and the Nonlinear Matter Power Spectrum}

In order to give an explicit demonstration of the approach, we first
generate a synthetic observational dataset from simulations. The key
advantage of doing this is that the underlying set of cosmological
parameters are known, allowing a direct test of the statistical
procedure. To generate the ``observations'' we begin with a smooth
power spectrum computed by running ten realizations of the same
cosmology with the parallel particle mesh (PM) code MC$^2$
(Cf. Heitmann et al. 2005 for code information and comparison results)
and averaging over the results. The initial conditions are set using
CMBFAST~\cite{cmbfast}. We restrict our study to the linear and
quasi-linear regime relevant to large-scale structure surveys; the
force resolution of the PM-code accurately resolves the scales of
interest.

\begin{figure}
\includegraphics[totalheight=75mm,angle=270,clip=]{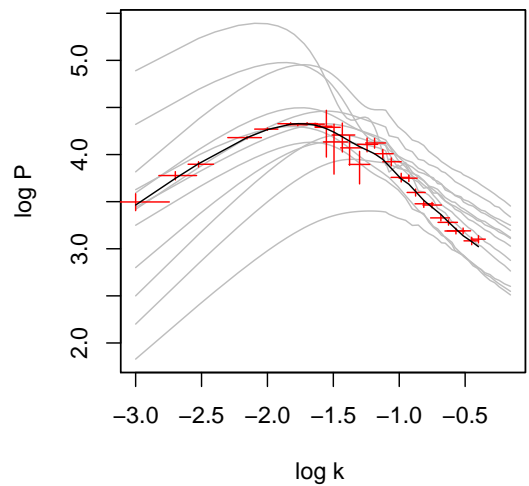}
\caption{Subset of the 128 simulated power spectra and the
synthetic dataset. The black line is the spectrum from which
the synthetic data were derived.}  
\label{plotone}
\end{figure}

We form a single power spectrum by attaching the linear $P_L(k),\;
0.001 h$Mpc$^{-1}\leq k \leq 0.1h$Mpc$^{-1}$ (growth specified by
linear theory), to the power spectrum from simulations $P_N(k)$ at $k
= 0.1h$Mpc$^{-1}$. Next, 28 points from this combined power spectrum
are picked, spaced roughly in the same bins as in a real dataset. The
error bars are set by values typical for cosmic microwave background
(CMB) experiments such as WMAP~(Sper\-gel et al. 2006) in the low-$k$
range, transitioning to values typical of surveys such as
SDSS~(Adelman-McCarthy et al. 2006) at higher $k$. Finally, points are
moved off the base power spectrum according to a Gaussian distribution
with a 1-sigma confidence, as shown in Figure~\ref{plotone}. Note that
for this test demonstration we are assuming that galaxy bias has
already been incorporated in the measurement. In a more realistic
situation, the bias would be included as part of the modeling
process. Note also that the choice of a homogeneous observational
dataset here is merely for convenience. For a heterogeneous dataset
such as CMB $C_l$'s combined with $P(k)$ measured from the galaxy
distribution, $y_{obs}$ [Cf. Eqn.  (1)] would also contain the CMB
results and the simulations underlying $\eta(\theta)$ would include
runs of (say) CMBFAST.

We consider the following five cosmological parameters:
$\theta=(n,h,\sigma_8,\Omega_{\rm CDM},\Omega_{\rm b})$. We assume a
flat $\Lambda$CDM universe with $\theta=(0.99,0.71,0.84,0.27,0.044)$
to make the synthetic observations (black line in
Figure~\ref{plotone}).  To determine the simulation design we must fix
the range that the input parameters should be varied over.  To this
end, we assume independent, flat priors over the ranges: $0.8\le n \le
1.4$, $0.5\le h \le 1.1$, $0.6 \le \sigma_8 \le 1.6$,
$0.05\le\Omega_{\rm CDM}\le 0.6$, and $0.02\le \Omega_{\rm b}\le
0.12$.  The simulation design prescribes a set of 128 input settings.
This number of simulations, as we show below, yields an emulator with
performance at the few percent level, sufficient for our present
purposes.

Each run is carried out with 128$^3$ particles on a 512$^3$ grid for a
450$h^{-1}$Mpc box, guaranteeing sufficient force resolution for the
scales of interest. To limit systematic biases, different seeds are
used to generate the initial Gaussian random field for each
simulation. Cosmic variance is minimized by matching the numerical to
the linear power spectrum in the linear regime near the $\Lambda$CDM
power spectrum peak. We show a subset of the 128 power spectra in
Figure~\ref{plotone}. The emulator is now built as described above --
note that the emulator is called only as needed by the MCMC analysis
in the calibration process.

\section{Results}

\begin{figure}[t]
\includegraphics[width=90mm]{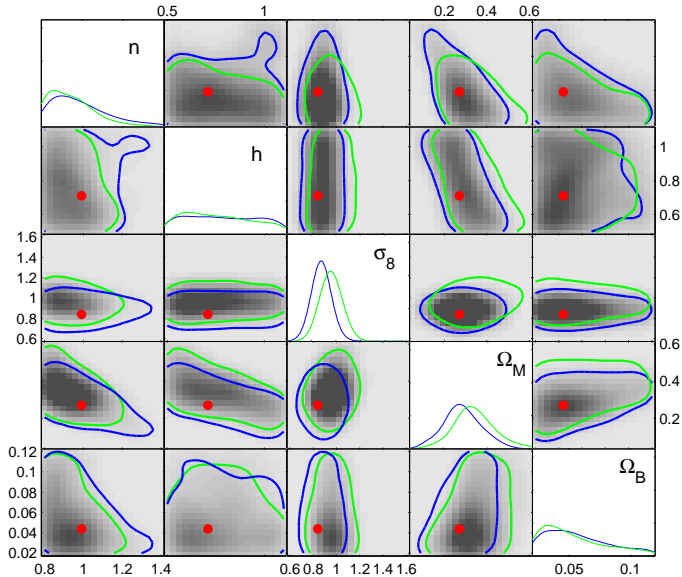}
\caption{Posterior density for the parameter vector $\theta$. 
The diagonal gives estimates of the univariate marginal pdfs for each 
component; blue: results from the entire synthetic dataset; green
using only the linear regime ($k < 0.1 h{\rm Mpc}^{-1}$). 
Off-diagonal
images show estimates of the bivariate marginal pdfs: upper triangle
for the entire dataset, lower triangle for the linear regime.  The
lines give  estimates of the 90\% highest posterior density
region. Again, blue is from using the entire dataset; green from using
only the linear regime. 
The dots show the actual parameter values used to generate
the synthetic observations.}
\label{plottwo}
\end{figure}

The posterior distribution of the five cosmological parameters is
depicted in Figure~\ref{plottwo}. The diagonal displays the
univariate, marginal pdfs for each of the parameters, while the
off-diagonal plots show estimated 2-d marginal densities, along with
90\% probability contours.  For comparison, Table~\ref{constraints}
gives the mean value of the parameters along with the estimated
uncertainty, as well as the ``true'' value for each parameter.  These
posterior estimates are obtained under two separate formulations --
one which uses all of the synthetic observation data, and one which
uses only the observations in the {\em linear} regime for which
$k<0.1h$Mpc$^{-1}$.  The green pdfs and contours in
Figure~\ref{plottwo} show the posterior results including information
only from the linear regime, whereas the blue pdfs and the contours
result from an analysis of the full nonlinear power spectrum.  Because
of the limited observational dynamic range, using only the linear
regime results in systematic shifts from the ``true'' answers, albeit
within the quoted uncertainties. Overall, we find $\Omega_{\rm CDM}$
and $\sigma_8$ to be very well determined. The full nonlinear analysis
over the entire $k$-range significantly improves the accuracy for
$\sigma_8$ and $\Omega_{\rm CDM}$ as well as the precision of the
constraint for $\sigma_8$ (see Table~\ref{constraints}). While the
synthetic dataset provides information about the remaining three
parameters, $n$, $h$, and $\Omega_{\rm b}$, they are not as well
constrained as is to be expected from an analysis restricted to the matter
power spectrum only. Note that the linear analysis underestimates the
uncertainty in $n$.

\begin{figure}
\includegraphics[width=45mm,angle=270]{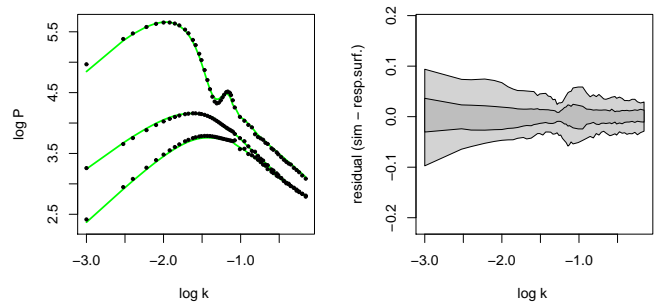}
\caption{Evaluation of the emulator fit.
Left: Three simulations (black dots) and the corresponding
response surface fits (green lines) obtained after holding out
the simulation to be predicted and training the response surface on
the remaining 127 simulations. Right: Residual (simulation $\log P - $
response surface) from holdout predictions (i.e. the simulation being
predicted is not used to estimate the response surface).  The central
gray region contains the middle 50\% of the residuals; the wider light
gray region, the middle 90\%.}      
\label{plotthree}
\end{figure}

\begin{table}
\begin{center}
\caption{\label{constraints}Parameter Constraints}
\begin{tabular}{lccc}
\tableline\tableline
Param.             &  Mean$^{\rm nonlin}$ &   Mean$^{\rm lin}$ &  True Value\\
\tableline
n                  &  $0.991^{+0.276}_{-0.171}$  &  $0.940^{+0.218}_{-0.132}$  &  0.99 \\
h                  &  $0.786^{+0.2823}_{-0.259}$  &  $0.765^{+0.287}_{-0.232}$  &  0.71 \\
$\sigma_8$         &  $0.882^{+0.082}_{-0.077}$  &  $0.962^{+0.121}_{-0.108}$   &  0.84 \\
$\Omega_{\rm CDM}$ &  $0.287^{+0.138}_{-0.133}$  &  $0.343^{+0.156}_{-0.130}$  &  0.27 \\
$\Omega_{\rm b}$   &  $0.057^{+0.052}_{-0.034}$  &  $0.054^{+0.052}_{-0.031}$  &  0.044 \\
\tableline\tableline
\vspace{-1.5cm}
\tablecomments{Mean value for the full and linear ($k<0.1h$Mpc$^{-1}$)
datasets with\\ their 90\% intervals, and the true value for the five
parameters under investigation.} 
\end{tabular}
\end{center}
\end{table}

\begin{figure*}[t]
\includegraphics[width=180mm]{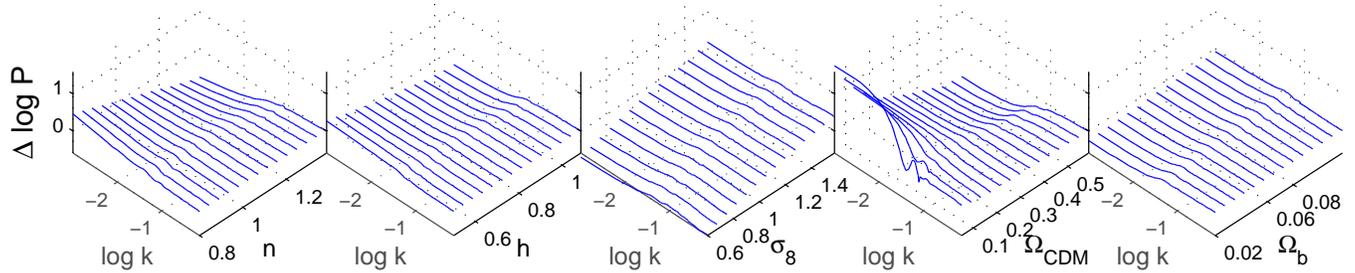}
\caption{Sensitivity of the computed power spectrum $\log P$ to
changes in input parameters.  Here, the response surface is used to
compute the change in $\log P$ as each parameter, in turn, is varied
from its lower bound to its upper bound while the other parameters
are held at their midpoints.}    
\label{plotfour}
\end{figure*}

The posterior distribution describes the uncertainty regarding the
parameter vector $\theta$ as well as statistical variance and
correlation parameters that control the response surface model.  Once
these posterior samples have been produced, it is straightforward to
generate posterior realizations of the emulator to assess its adequacy
in modeling the simulated output.  The accuracy of the emulator was
estimated by excluding individual simulation runs and building a new
emulator based on the remaining 127 power spectra. The emulator
predictions can now be compared against the actual simulation output
of the excluded run. Three examples of applying this procedure are
shown in the left plot in Figure~\ref{plotthree}. The accuracy of the
emulator turns out to be extremely good, at the level of a few
percent, which is very adequate for the present analysis. The right
panel in Figure~\ref{plotthree} summarizes the residuals for all 128
simulations -- the central gray band delineates the middle 50\% of the
residuals; the light gray band delineates the middle 90\%.  Gaussian
process models offer a number of advantages over other methods for
modeling simulation output: they do not require runs over a grid of
input settings; they allow for interpolation of the simulation output;
they can accommodate fairly general interactions between input
parameters; and typically outperform other modeling approaches.  For
example, the GP model gives substantially better predictions as
compared to a quadratic response surface model, a generalized additive
model (GAM), or a multivariate additive regression spline model (MARS)
(Hastie et al. 2001).

The fitted emulator can be used to explore the sensitivity of the
simulation output to changes in the cosmological
parameters. Figure~\ref{plotfour} shows how the log of the power
spectrum changes as one parameter is varied, the others being fixed at
their prior midpoints. Both $\sigma_8$ and $\Omega_{\rm CDM}$ have a
large impact when varied over their prior ranges.  Hence it is not
surprising that the posterior distribution for these two parameters
are the most constrained by the observed data.  Figure~\ref{plotfour}
also suggests that while most parameters affect the power spectrum in
the linear regime ($k < 0.1h{\rm Mpc}^{-1}$), only $\sigma_8$ affects
the power spectrum in the nonlinear regime ($k > 0.1h{\rm
Mpc}^{-1}$). Thus, while additional data in the nonlinear regime is
likely to help constrain $\sigma_8$, it will not greatly reduce
uncertainty in the other four parameters.

\section{Conclusion and Outlook} 

We have introduced a new, very powerful method for determining
cosmological and model parameters from simulations and
observations. The key idea is to extract maximum utility from a
necessarily finite set of expensive simulations. The implementation of
this idea includes several valuable features: (i) a design to
optimally sample the simulation parameter space; (ii) an accurate
emulator capable of generating the required outputs in between the
sampled simulation points; (iii) an uncertainty and sensitivity
analysis; (iv) the parameter constraints themselves, with associated
uncertainty bounds. 

In order to demonstrate the basic approach, we used a set of 128 dark
matter structure formation simulations and a homogeneous synthetic
``observational'' dataset to determine five cosmological
parameters. The next step is to use the framework for analyses of real
data, especially of combined datasets such as the CMB and large scale
structure observations.

There are many ways to enhance the method and improve its
performance. One is the melding of information from codes with
different degrees of resolution and input physics, such as in the
extraction of information about the mass distribution from the
Lyman-$\alpha$ forest. Here, complex hydrodynamics simulations are
certainly desirable, but much faster approximate methods such as
hydro-particle mesh (HPM) are available. Thus, a first analysis based
on HPM can be performed, narrowing the parameter range of interest
sufficiently to make hydro runs feasible. Interesting offshoots of the
methodology include the exploitation of certain intermediate
results. For instance, a large set of N-body simulations can be
performed with several input parameters such as the equation of state
for dark energy. An emulator can then be constructed from these and
publicly released. This emulator can then be conveniently used instead
of real simulations for planning observations and data analysis.

\acknowledgements 
We thank Brian Williams for creating the simulation designs and Kevork
Abazajian, Lam Hui, and Adam Lidz for useful discussions and
encouragement. A special acknowledgment is due to supercomputing time
awarded to us under the LANL Institutional Computing Initiative. This
research is supported by the DOE under contract W-7405-ENG-36.

\end{document}